\documentclass[%
 reprint,
 amsmath,amssymb,
 aps,
floatfix,
]{revtex4-2}

\usepackage[dvipdfmx]{graphicx}
\usepackage{dcolumn}
\usepackage{bm}
\usepackage[pdfpagelabels]{hyperref}

\usepackage{here} 
\usepackage[raggedright]{subfigure} 
\usepackage{color}
\numberwithin{equation}{section} 

\newcommand{\unit}[1]{\ \mathrm{#1}}

\begin{document}

\preprint{APS/123-QED}

\title{Application of the Hilbert-Huang transform for analyzing standing-accretion-shock-instability induced gravitational waves in a core-collapse supernova}

\author{M. Takeda$^{1}$} 
\author{Y.~Hiranuma$^{2}$}
\author{N.~Kanda$^{1,3}$}
\author{K.~Kotake$^{4,5}$} 
\author{T.~Kuroda$^{6}$}
\author{R.~Negishi$^{2}$}
\author{K.~Oohara$^{2,7}$}
\author{K.~Sakai$^{8}$} 
\author{Y.~Sakai$^{9}$}
\author{T.~Sawada$^{1}$}
\author{H.~Takahashi$^{7,9,10}$}
\author{S.~Tsuchida$^{1}$} 
\author{Y.~Watanabe$^{2}$}
\author{T.~Yokozawa$^{11}$}
\affiliation{
$^{1}$Graduate School of Science, Osaka City University, Sumiyoshi-ku, Osaka City, Osaka 558-8585, Japan\\
$^{2}$Graduate School of Science and Technology, Niigata University, Nishi-ku, Niigata City, Niigata 950-2181, Japan\\
$^{3}$Nambu Yoichiro Institute of Theoretical and Experimental Physics (NITEP), Osaka City University, Sumiyoshi-ku, Osaka City, Osaka 558-8585, Japan\\
$^{4}$Department of Applied Physics, Fukuoka University, Jonan-ku, Fukuoka City, Fukuoka 814-0180, Japan\\
$^{5}$Research Institute of Stellar Explosive Phenomena, Fukuoka University, Jonan-ku, Fukuoka City, Fukuoka 814-0180, Japan\\
$^{6}$Max-Planck-Institut f{\"u}r Gravitationsphysik, D-14476 Potsdam-Golm, Germany\\
$^{7}$Institute for Cosmic Ray Research, The University of Tokyo, Kashiwa City, Chiba 277-8582, Japan\\
$^{8}$Department of Electronic Control Engineering, National Institute of Technology, Nagaoka College, Nagaoka City, Niigata 940-8532, Japan\\
$^{9}$Research Center for Space Science, Advanced Research Laboratories and Graduate School of Integrative Science and Engineering, Tokyo City University, Setagaya-ku, Tokyo 158-0082, Japan \\
$^{10}$ Earthquake Research Institute, The University of Tokyo, Bunkyo-ku, Tokyo 113-0032, Japan\\
$^{11}$Institute for Cosmic Ray Research, KAGRA Observatory, The University of Tokyo, Hida City, Gifu 506-1205, Japan
}


\begin{abstract}
Through numerical simulations, it is predicted that the gravitational waves (GWs) reflect the characteristics of the core-collapse supernova (CCSN) explosion mechanism. There are multiple GW excitation processes that occur inside a star before its explosion, and it is suggested that the GWs originating from the CCSN have a mode for each excitation process in terms of time-frequency representation. Therefore, we propose an application of the Hilbert-Huang Transform (HHT), which is a high-resolution time-frequency analysis method, to analyze these GW modes for theoretically probing and increasing our understanding of the explosion mechanism. The HHT defines frequency as a function of time, and is not bound by the trade-off between time and frequency resolutions. In this study, we analyze a gravitational waveform obtained from a three-dimensional general-relativistic CCSN model that showed a vigorous activity of the standing-accretion-shock-instability (SASI). 
We succeed in extracting the SASI induced GWs with high resolution on a time-frequency representation using the HHT and we examine their instantaneous frequencies. 
\end{abstract}

\maketitle


\section{\label{sec:level1}Introduction}


Advanced LIGO~\cite{aasi2015advanced} and
Advanced Virgo~\cite{acernese2014advanced} detected 11 gravitational
wave (GW) events during the first and the second observing run (O1 and O2)
~\cite{abbott2019gwtc}. The third observing run (O3) started in April
2019 but was globally suspended in March 2020. It has been reported
that 39 candidate GW events were
found in the data during the first half of O3~\cite{abbott2020gwtc2}.
GEO600~\cite{Grote_2004} and
KAGRA~\cite{akutsu2018kagra, akutsu2020overview}
conducted a joint two-week observing run (O3GK) in April 2020.
The next fourth observing run (O4) will be a joint observation of
LIGO, Virgo and KAGRA.

The only type of event currently detected is a compact binary coalescence (CBC), 
and new sources, such as the core-collapse supernovae (CCSNe), are expected to be detected by using the third-generation GW detectors 
(e.g., Einstein Telescope~\cite{punturo2010einstein}, 
Cosmic Explorer~\cite{abbott2017exploring} etc.)~\cite{kalogera2019yet}.

The GWs from CBC are generally analyzed  by employing a matched filter (MF) that uses model waveforms. 
However, a proper model waveform can hardly be created for the GWs from CCSNe (see 
\cite{abdikamalov,kotake13} for a review) and therefore, it is in general difficult to apply the MF in this case.
These unmodeled GWs are usually analyzed by
some types of time-frequency representation (TFR) techniques.


Since the GWs from CCSNe have characteristic modes in TFR, spectrogram analysis has recently become the mainstream method for detecting them and classifying CCSNe models, rather than analyzing time-series waveforms~\cite{klimenko2008coherent, klimenko2016method, gossan2016observing, logue2012inferring, powell2016inferring, powell2017inferring, roma2019astrophysics, suvorova2019reconstructing, astone2018new, iess2020core, chan2020detection}.
In particular, the mainstream detection method is TFR analysis of coherent multidetector networks in the wavelet domain~\cite{klimenko2008coherent, klimenko2016method, gossan2016observing}.
Principal component analysis has also been shown to be useful for detection and classification, and software for supernova classification has been developed~\cite{logue2012inferring, powell2016inferring, powell2017inferring, roma2019astrophysics, suvorova2019reconstructing}.
Furthermore, machine learning has been shown to be useful for detection and classification, and these methods are already capable of properly analyzing signals embedded in the non-Gaussian and nonstationary noise of the GW detectors~\cite{astone2018new, iess2020core, chan2020detection, lopez2021deep}.
This paper proposes a new application of TFR with the ultimate goal of analyzing the structure of detected and classified GWs from the CCSNe in more detail.



The importance of GWs in the CCSNe research is demonstrated by the results of numerical simulations. 
It has been hypothesized that the GWs carry information about the star's internal state before its explosion 
and consists of the GWs from several emission mechanisms that occurred in the star’s core. 
Although there are several multidimensional general relativistic hydrodynamics simulations to obtain, at least, 
a theoretical explanation of the mechanism of CCSNe, it is not yet fully understood. 
The characteristics of the GW waveform depends on the emission mechanism, and GWs could be the ``smoking gun'' 
for the phenomena that take place inside the star.
However, the theoretically predicted GW amplitudes from CCSNe are extremely small and are usually hidden by the background noise~\cite{gossan2016observing}.
Therefore, together with the poorly understood CCSN explosion mechanisms, it makes it difficult to even detect the GW signals in the expected real events, and even more to decipher their physical meanings to infer the phenomena occurring inside the stellar core.

In the event of a gravitational collapse, the core temperature and density of the collapsed star increases rapidly, thereby trapping the neutrinos. 
Following this, when the density exceeds the nuclear saturation density, the core rapidly regains pressure, which competes 
with its self-gravity and causes core bounce. 
This leads to the formation of the shock wave in the vicinity of the inner core. At the core bounce, the unshocked core also gives rise to the formation of a protoneutron star (PNS),  the evolution of which depends on the explosion dynamics (e.g., mass accretion rate onto it) in the postbounce phase.  The bounce shock does not immediately 
blow off the outer layer of the collapsed core, but is temporarily stalled (e.g.,~\cite{kotake06} for a review). Before the possible onset of the neutrino-driven explosions,  the stalled shock is known to be susceptible to multidimensional fluid instabilities (see \cite{jankarev,burrowsrev,foglizzorev}  for recent reviews) including neutrino-driven convection and the standing accretion shock instability (SASI \cite{blondin03}), the latter of which globally deforms the shock surface.
The briskness of the SASI depends on the nuclear equation of state (EoS) and initial conditions for the simulation, such as progenitor compactness and rotation.
One of its remarkable features related to the GW emission mechanism is that, once it is fully grown, it produces a time modulated mass accretion onto the PNS core.
The time modulated mass accretion penetrates deep into the PNS and triggers a surface distortion and motion of the central PNS core.
From there and also from the globally deformed regions behind the shock \cite{tonygw}, sizeable GWs could be emitted with their time frequency being characterized by the SASI frequency of the order of $\sim100$ Hz \cite{kuroda2016sasi,Andresen2017gravitational,burrowsgw,tonygw,shibagakigw,powellgw,Vartanyan_2020,10.1093/mnras/stz2307}
(we call them SASI mode hereafter).

The models are based on numerical-relativity simulations taking into account a gray energy neutrino transport.
In this paper, we particularly focus on the GWs emitted from the SASI activities by discussing models developed by Kuroda \textit{et al.}~\cite{kuroda2016sasi}.
The models are based on numerical-relativity simulations taking into account a gray energy neutrino transport.
In fact, the GW simulated by these previous studies has the SASI mode in the time-frequency representation and its frequency band ($\sim100-200 \unit{Hz}$) is significantly smaller than the PNS surface g-mode oscillation ($\sim 700 \unit{Hz}$) induced by convection.
The details of the behavior of the PNS in the simulation when the SASI occurs have been investigated by Kawahara \textit{et al.}~\cite{kawahara2018linear} using GW analysis.

We further suggest that Hilbert-Huang Transform (HHT)~\cite{huang1996mechanism,huang1998empirical,huang1999new,huang2014hilbert} can be used to investigate whether we can actually capture the onset of the SASI, and if so, to what extent we can capture the nature of the SASI.

HHT was first applied to the analysis of 
GWs in 2007~\cite{camp2007application}.
Here, we highlight two advantages of using the HHT for analyzing the GWs.
First, the HHT does not have the time and frequency resolution trade-off because it defines frequency as a function of time. 
This is called the instantaneous frequency (IF).
Moreover, the amplitude, phase, and frequency acquired by the HHT have the same time resolution as the original time series signal.
Second, the HHT decomposes the original time series signal into several time series signals that oscillate around the zero point.
This process should be able to separate and obtain those modes from the GWs in which multiple radiation processes are mixed. 
Hence, it is fit for the analysis of GWs from the CCSNe.
The effectiveness of the HHT in the GW analysis has been shown by studies assuming binary black hole, binary neutron star, 
and burst~\cite{stroeer2009methods,takahashi2013investigating,nakano2019comparison,kaneyama2016analysis,kaneyama2017research}.
Furthermore, the HHT also worked well for the analysis of GW150914, which is the first GW detection event~\cite{sakai2017estimation}.


This paper proposes a method with HHT of analyzing gravitational waves from a core collapse supernova.
This is important to decipher what could be the driving mechanism of supernova explosion from observed GWs in the future.
We have reanalyzed a GW analyzed by~\cite{kuroda2016sasi,kawahara2018linear} with the HHT.
In particular, we focused on how clearly and accurately the SASI mode can be extracted and showed the HHT is a promising tool for it.
Detailed analyses of results of various simulations will be given elsewhere.

This paper is organized as follows: 
In Sec.~\ref{sec:level2}, we introduce an overview of the HHT.
In Sec.~\ref{sec:3DCCSNmodel}, we describe the dynamics of a 3D CCSN model of our interest and its GW emission mechanisms.
In Sec.~\ref{sec:level3}, we demonstrate the time-frequency representation using the HHT.
In Sec.~\ref{sec:level4}, we propose our method of analyzing the frequency and apply it to test the signal and simulated GWs. 
Finally, we summarize and discuss our results in Sec.~\ref{sec:level5}.


\section{\label{sec:level2}Hilbert-Huang Transform}

The HHT consists of two steps. The first step is the empirical mode
decomposition (EMD) that decomposes the original signal into
intrinsic mode functions (IMFs). The second step is the
Hilbert spectral analysis (HSA) of each IMF~\cite{huang1998empirical}.

The HSA is a type of TFR using the Hilbert transform (HT), 
\begin{align}
  \mathcal{H}[u(t)]
  = \frac{1}{\pi}
  \mathop{\mathrm{PV}} \int^{+\infty}_{-\infty} \frac{u(\tau)}{t-\tau} d\tau
  = u(t) \ast \left( \frac{1}{\pi t} \right),
  \label{eq:HT}
\end{align}
where $u(t)$ is a real function of time, 
$\mathrm{PV}$ and $\ast$ denote the Cauchy principal value of the singular integral and convolution, respectively.
Using the HT, the complex signal $F(t)$ can be defined
from a real signal $u(t)$~\cite{cohen1995time} as
\begin{equation}
  F(t) = u(t) + i \mathcal{H}[u(t)]
  = a(t) \exp \big[ i \phi(t) \big],
\end{equation}
in which
\begin{align}
  a(t) &= \sqrt{u(t)^2 + \mathcal{H}[u(t)]^2},\\
  \phi(t) &= \arctan \left( \frac{\mathcal{H}[u(t)]}{u(t)}\right).
\end{align}
Note that if $u(t)$ is the real part on the real axis of
a complex function $F(z)$ that is analytic in the upper half complex plane and
goes to zero rapidly enough for $|z| \rightarrow \infty$,
the imaginary part of $F(z)$ on the
real axis is uniquely given by the HT of $u(t)$, provided the HT
in Eq.(\ref{eq:HT}) exists.
Here,
$a(t)$ and $\phi(t)$ can be considered as the instantaneous amplitude (IA) and the
instantaneous phase of the signal, respectively,
if the characteristic time scale of variation of $a(t)$ is longer than
that of $\phi(t)$. The instantaneous frequency (IF) is simply given by
\begin{equation}
  f(t) = \frac{1}{2\pi} \frac{d\phi(t)}{dt}.
\end{equation}

The HSA gives IA and IF which have a clear physical picture with a
signal of a single oscillation mode. However, it is not the case with
a signal including multiple oscillation modes.  
Huang \textit{et al.}~\cite{huang1998empirical} proposed that this fault can be overcome
if an actual observed signal $s(t)$ is decomposed into a finite number of IMFs
$c_j(t)$ along with a residual $r(t)$ as
\begin{equation}
  s(t) = \sum_{j=1}^{N} c_j(t) + r(t),
\end{equation}
by using the EMD.  Here each IMF $c_j(t)$ oscillates around zero and
$r(t)$ is a nonoscillation mode.  The EMD is a series of adaptive
sifting processes that reveal high-frequency impacts without leaving
the time domain. The first IMF (IMF1), $c_1(t)$, has the shortest
characteristic timescale, that is, it represents the highest
frequency mode of the original signal.

The HHT is very useful to analyze nonstationary and nonlinear signals,
but the simple EMD encounters weighty problems in some
cases. Wu and Huang~\cite{wu2009ensemble} introduced the ensemble EMD
(EEMD) to solve the mode-mixing problem.

The EEMD consists of the following steps:
\begin{enumerate}
\item Add a Gaussian white noise series to the original signal.
\item Apply the EMD to the data with added white noise to obtain IMFs.
\item Make ensembles of IMFs by repeating step 1 and step 2 many times
  with different white noise series added each time.
\item Take the ensemble mean of each ensemble to obtain the final
  result of each IMF.
\end{enumerate}
In these steps, the mode mixing is effectively eliminated, while the
added noise is averaged out through ensemble average. There are two
significant parameters that need to be predetermined in these
steps. The one is the amplitude of added noise.  It depends on the
characteristics of the original signal but it is usually suggested
that the ratio of the standard deviation of the Gaussian white noise
to that of the original signal, $\sigma_{\mathrm{eemd}}$, is less than
unity. However, Hiranuma \textit{et al.}~\cite{hiranuma2021eemd} found that the larger value is
appropriate to effectively eliminate the mode mixing with the signal
that will be analyzed in this paper. The other parameter is the number
of ensemble trials $N_{\mathrm{eemd}}$, which mainly depends on the
value of $\sigma_{\mathrm{eemd}}$. Although the value of
$N_{\mathrm{eemd}}$ is infinity to cancel out the effect of the added
withe noise completely, too large value would increase computational
cost. After some trials of these parameters with our signal, we found
that the most appropriate value of $\sigma_{\mathrm{eemd}}$ is 10.
For this amplitude, a very large value of
$N_{\mathrm{eemd}}$ is required. It was found that the final results
do not change with the number of trial larger than $10^6$ at most.
Hence we set $\sigma_{\mathrm{eemd}}$ and $N_{\mathrm{eemd}}$
on 10 and $10^6$, respectively, in this paper unless otherwise indicated.
We used KAGRA Algorithmic Library (KAGALI)~\cite{kagali} for HHT calculation.

\section{\label{sec:3DCCSNmodel}3D CCSN model}

In this section, we describe the dynamics of a 3D CCSN model of our interest and its GW emission mechanisms before explaining the applicability of HHT analysis for simulated gravitational waveform.
The initial conditions are \citep{kuroda2016sasi}: the progenitor mass is $15 M_{\odot}$, no rotation, and the EoS is ``SFHx''.
This simulation was calculated up to 0.35 s from the core bounce time, and 
its GW was derived using the standard quadrupole formula, assuming a distance of $10$ kpc.

Shortly after core bounce, the initial post bounce convection develops due to the negative entropy gradient.
It emits GWs at a relatively low frequency range of $\sim100$ Hz and lasts till $\sim100$ ms after bounce.
Afterward the dominant hydrodynamic activity can be taken over by a nonlinear SASI motion, while the PNS convection still coexists.
Once the SASI phase initiates, the standing shock surface globally deforms its morphology depending on the dominant SASI mode, which varies from moment to moment.
For instance in the current 3D CCSN model, the dominant mode is initially a sloshing mode, which gradually shifts to a rotating spiral mode around $200$ ms after bounce.
In either mode, a typical frequency of SASI appears at $\sim100$ Hz, which can be derived from shock radii $\sim100$ km divided by the advection speed of fluid in the shocked region $\sim10^8$ cm s$^{-1}$.
At $\sim300$ ms after bounce, the hydrodynamic activity is again dominated by the convective motion.

Based on the 3D CCSN model, the ahthors of \cite{kuroda2016sasi} discussed a new gravitational wave emission mechanism in association with the vigorous SASI activities.
They demonstrated that the low-frequency GW emissions can be triggered by a strong mass accretion having the same temporal dependence with the SASI frequency \citep[see also][]{Andresen2017gravitational,burrowsgw,Vartanyan_2020}.
In the next section, we will discuss how such a characteristic frequency, which usually varies with time, can be extracted from the waveform using the HHT.

Here we should comment on the neutrino treatment used in the current 3D CCSN model. 
It was calculated by a gray energy neutrino transport that is no longer state-of-the-art simulation compared to the recent spectral neutrino transport simulations.
We, however, anticipate that the main focus of this study, i.e., extracting the characteristic GW frequencies associated with the SASI, would not be affected even if we use another model with more sophisticated neutrino transport.
This is because the development of SASI and its GW emission do not sensitively depend on the neutrino transport method.
It can be justified from recent full-fledged 3D spectral neutrino transport simulations such as \cite{Andresen2017gravitational,powell2017inferring,Vartanyan_2020}, in which they reported the SASI developments.


\section{\label{sec:level3}Application of HHT for a simulated GW}

\begin{figure*}[th]
  \centering
  \subfigure[STFT]{\includegraphics[width=0.48\linewidth]{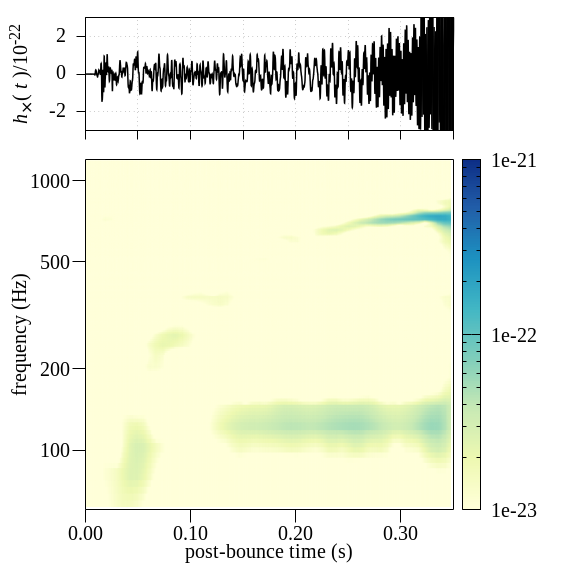}
    \label{fig:tfm_stft}}
  \hspace{0.01\linewidth}
  \subfigure[HHT]{\includegraphics[width=0.48\linewidth]{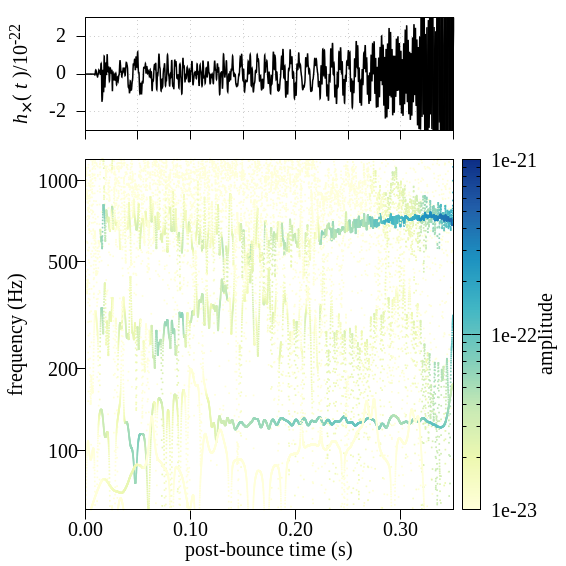}
    \label{fig:tfm_hht}}
  \caption{Time-Frequency representation by STFT spectrograms and HHT map. The right and left top panels show the same GW signal wave ($h_{\times}$) ``SFHx''~\cite{kuroda2016sasi} and sampling frequency of the signal is 16 kHz. In the bottom panels of (a) and (b), the $x$ axis and $y$ axis represent the post core bounce time and frequency in logarithmic scale, respectively. STFT's color bar denote amplitude spectrum and HHT's color bar denotes instantaneous amplitude. In the STFT calculation, we used the Hanning window function and adopted $\sim 0.048\unit{s}$ window width and $\sim 0.042\unit{s}$ overlap.}
		 \label{fig:tfmap}
	\end{figure*}

We show the applicability of TFR using the HHT for a GW signal from the aforementioned three-dimensional general-relativistic (GR) core-collapsed supernova simulation~\cite{kuroda2016sasi}.
Both upper panels in Fig.~\ref{fig:tfmap} show the same GW signal ($h_{\times}$) \cite{kuroda2016sasi} 
and the lower panels show the time-frequency map calculated using short time Fourier transform (STFT) [Fig.~\ref{fig:tfm_stft}] and the HHT [Fig.~\ref{fig:tfm_hht}].
There are two strong signals in both the time-frequency domains: 
one is the PNS surface g-mode (gravity-mode), whose frequency increases rapidly with time at about $0.05 - 0.35 \unit{s}$; 
the other has a lower frequency than the g-mode and is suggested to be induced by the SASI mode.
Using the STFT calculations, we can estimate that the SASI mode has a low frequency band and is approximately constant. 
We confirmed that there is no significant difference between results with
$h_{\times}$ and $h_+$, and therefore we show only the results with
$h_{\times}$ here and hereafter.


Hence, we compare our results with the previously reported outcomes~\cite{kuroda2016sasi,kawahara2018linear}.
Apart from the two strong modes (g-mode, SASI mode), there are several other modes in TFR,
also appear in the HHT. However, the IAs, which denote the weight of the wave with the frequency exist, are small.
Thus, we do not focus on these modes and analyze only the SASI mode.

	\begin{figure*}[ht]
          \centering
          \subfigure[IMFs of the GW using EEMD]{
            \includegraphics[width=0.48\linewidth]{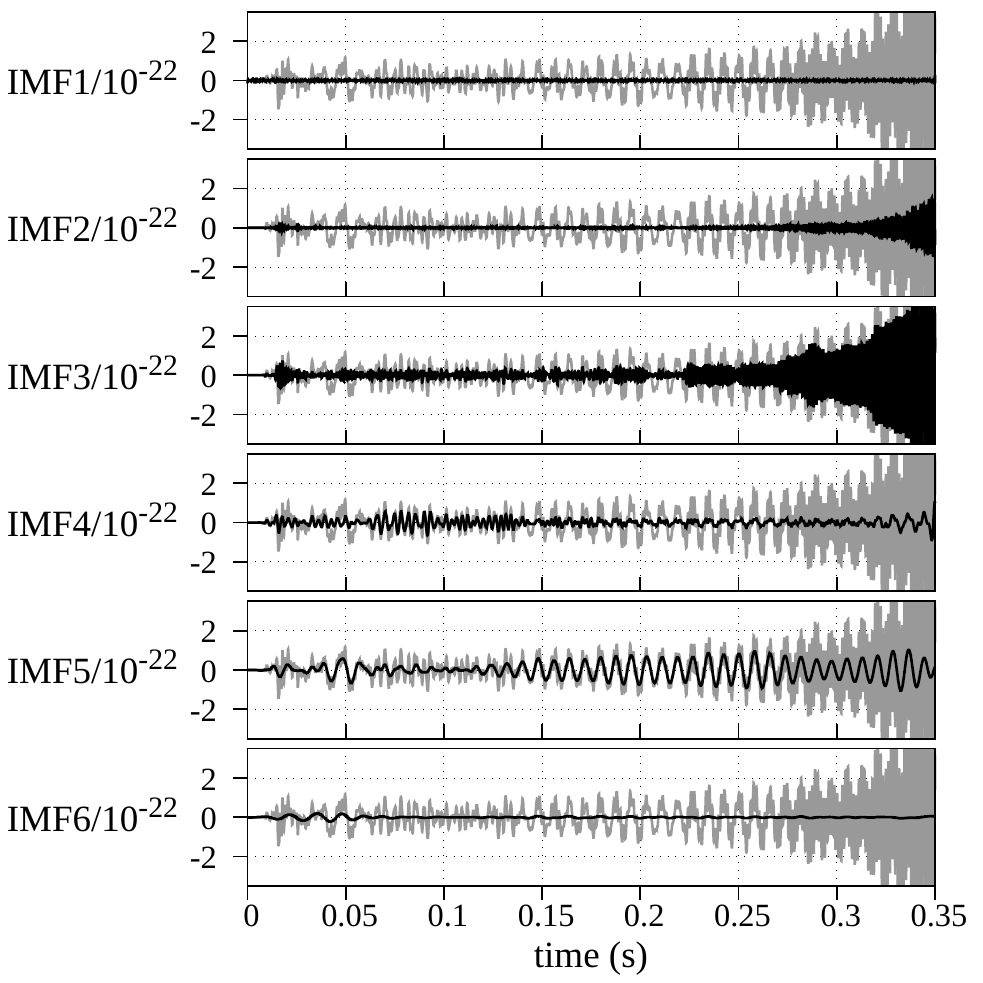}
            \label{fig:imfs}}
          \hfill
          \subfigure[IMF5, IA and IF of IMF5]{
            \includegraphics[width=0.48\linewidth]{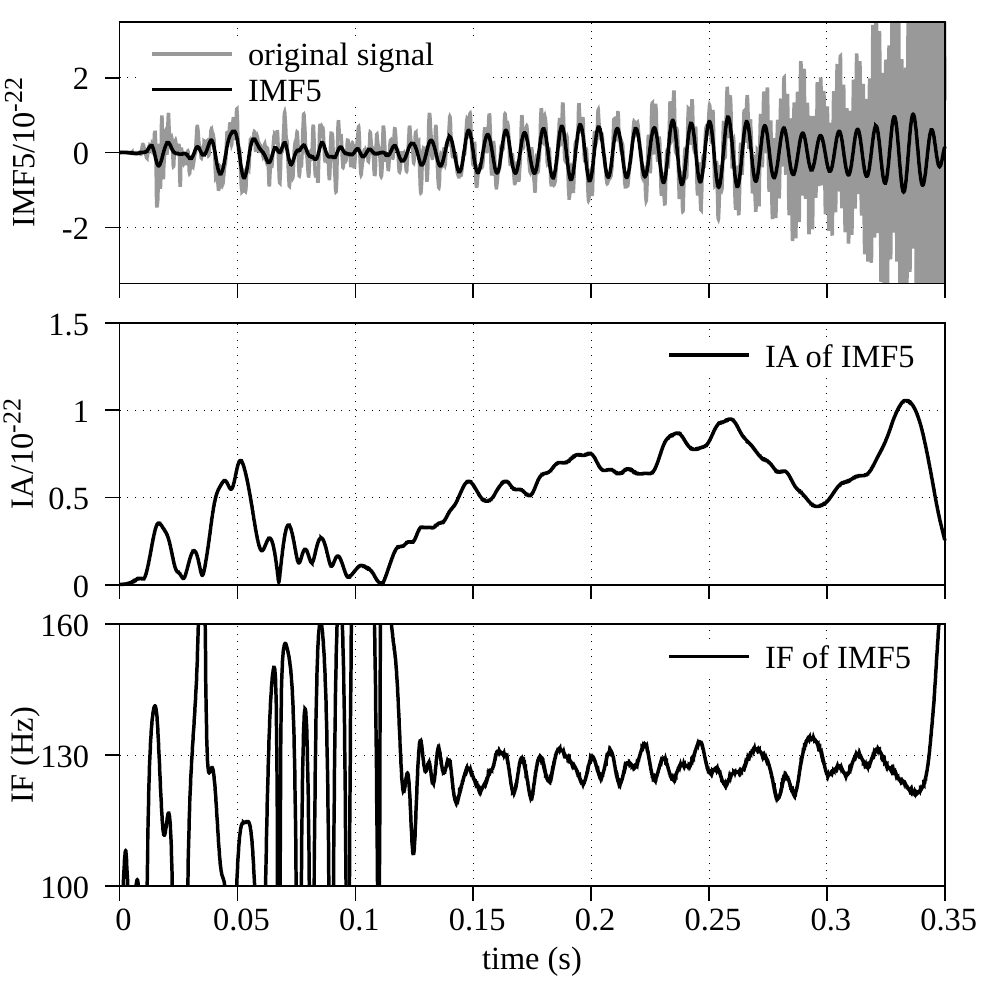}
            \label{fig:iaifimf5}}
          \caption{IMF1-6 by EEMD and the IA and IF of the IMF5 by HSA. The HHT map shown in Fig.~\ref{fig:tfm_hht} was created by HSA all of the IMFs in Fig.~\ref{fig:imfs}.}		
          \label{fig:imf}
	\end{figure*}

We decomposed the GW signal into the IMFs and showed the IMF1-6 in Fig.~\ref{fig:imfs}.
The g-mode and SASI mode components were extracted as the IMF3 and IMF5, respectively.
Fig.~\ref{fig:iaifimf5} illustrates the IMF5, its IA and IF, and indicates that the SASI mode did not appear up to about $ \sim 0.1 \unit{s}$ from the core bounce time and 
it perhaps oscillates with about $\sim 150 \unit{Hz}$ after that.

Since the HHT has a higher time and frequency resolution than that of the STFT, the HHT is suitable for identifying the SASI dominant phase and investigating the time dependency of its GW frequency.
Extracting the instantaneous characteristic GW frequency from detected GWs particularly exerts its ability in multimessenger astronomy.
This is because the SASI activity may imprint its time modulated signal in both the gravitational waveform \cite{kuroda2016sasi,Andresen2017gravitational,Vartanyan_2020} and also in the neutrino signal \cite{Tamborra13,kuroda2017correlated}.
We, therefore, need to determine more precisely the characteristic GW frequency to link the GW and neutrino signal and to identify that they are emitted from the same origin.
For that purpose, we propose a method to investigate the characteristics of a GW frequency from the SASI based on the HHT.

In Sec.~\ref{sec:level4}, we discuss how to analyze the SASI mode properly and show our results.

\section{\label{sec:level4}Analysis of the instantaneous frequency}
We show the method to analyze the IF of the IMF5, which includes the SASI mode.
The activity of the GWs derived from SASI changes depending on the selection of the mass of progenitors and EoS, and the theory that describes their properties has not been confirmed~\cite{kuroda2016sasi,kuroda2017correlated}.
We propose a method for investigating the characteristics of the GWs derived from SASI using the HHT.
Here, we analyze the IF to estimate the time when the SASI mode occurred, 
and investigate the average and trend of the IF. 
Furthermore, the systematic error discovered by applying the same method to the test waveform imitating the GW is described.

\begin{figure*}[ht]
  \centering
  \subfigure[]{
    \includegraphics[width=0.9\linewidth]{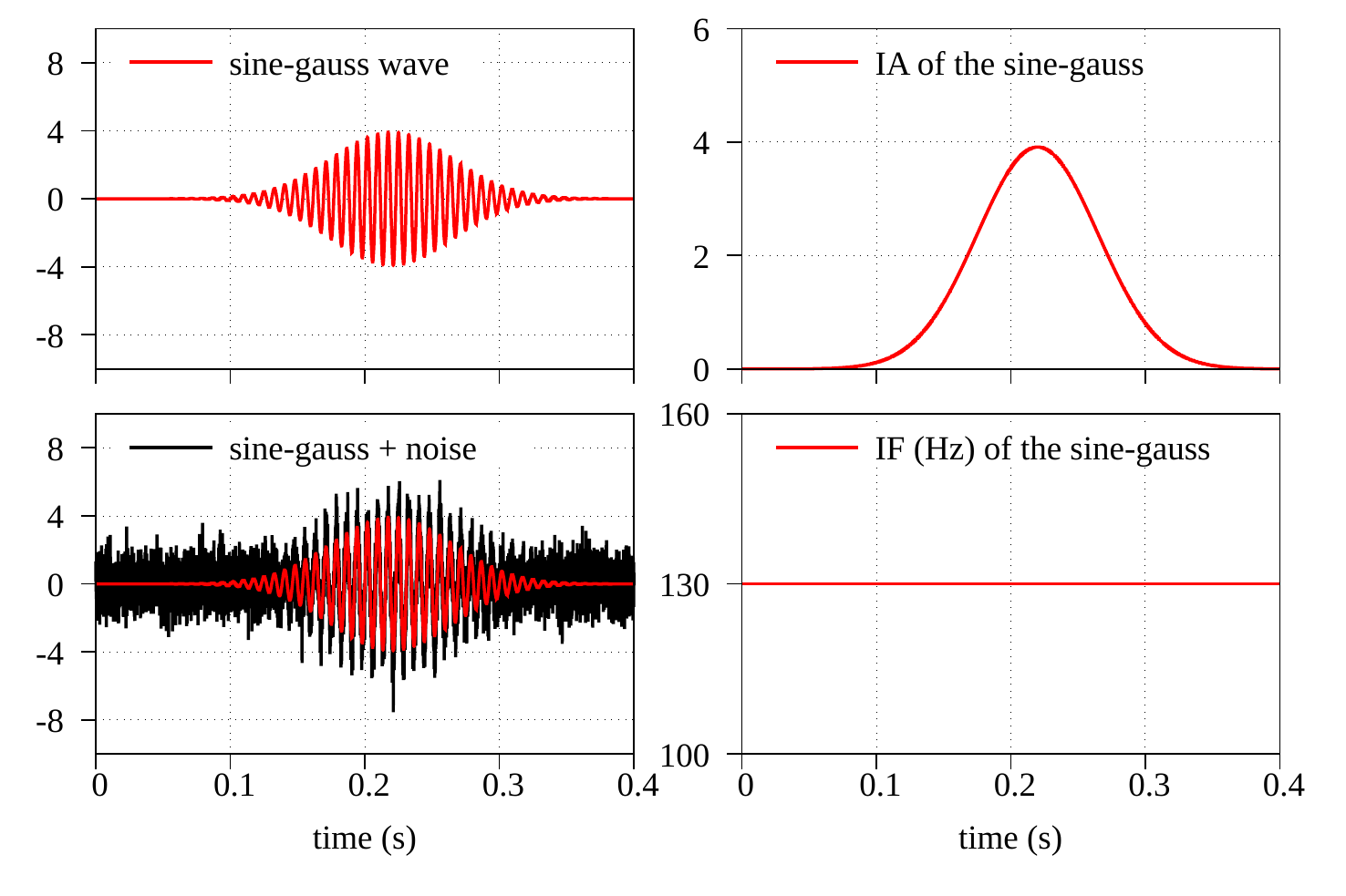}
    \label{fig:testsignal}}

  \subfigure[]{
    \includegraphics[width=0.47\linewidth]{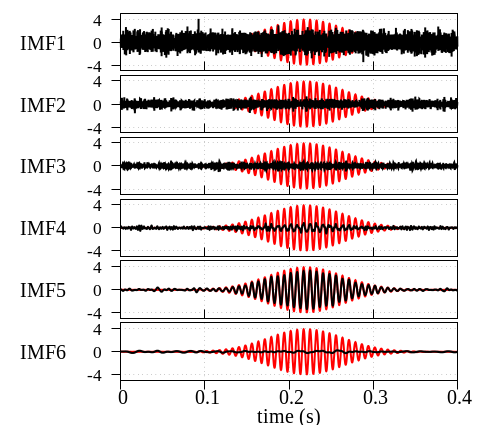}
    \label{fig:testimf}}
  \hspace{0.01\linewidth}
  \subfigure[]{
    \includegraphics[width=0.47\linewidth]{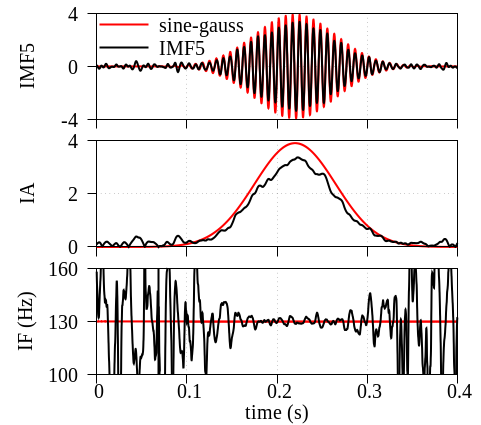}
    \label{fig:testhsa}}
  \caption{Results of analysis of the test signal shown in Eq.~\eqref{eq:testsignal} using HHT.
  (a) Test signal shown in Eq.~\eqref{eq:testsignal} 
  (b) IMFs of test signal \eqref{eq:testsignal} using EEMD
  (c) IMF5, IA and IF of test signal \eqref{eq:testsignal} using HSA.}
  \label{fig:testhht}
\end{figure*}	

\subsection{Method and demonstration}
We use a sine-Gaussian waveform with constant frequency as the test signal,
	\begin{align}
		  h_{\mathrm{test},l}(t) =& 
		  A 
		  \exp \left[ 
		  		- \left( 
					\frac{2 \pi (t-t_\mathrm {shift})}{t_\mathrm{width}} 
				\right)^2 
			    \right] 
		\sin (2 \pi f_{\mathrm{const}} t)\notag \\ &+ g_l,
 		 \label{eq:testsignal}
	\end{align}
where $g_l$ is the white Gaussian noise,
$l$ is $(l  = 1,\cdots,N_\mathrm{test})$,
$N_\text{test}$ is number of the white Gaussian noise signals generated from different seeds.
In Eq.~\eqref{eq:testsignal}, 
the former simulates the SASI mode, 
and the latter simulates the numerical error.
We used the parameters 
$t_\mathrm{shift}=0.22, 
t_\mathrm{width}=0.4, 
f_{\mathrm{const}} = 130 \unit{Hz}$ 
and the corresponding signal is shown in Fig.~\ref{fig:testsignal}.
The reason for using a sine-Gaussian wave is that, as shown in Fig.~\ref{fig:tfmap}, 
the SASI has a waveform in which the amplitude becomes stronger from a certain time, 
the frequency is almost constant, and then disappears after vibration.
The SASI is a phenomenon whose generation conditions and behavior are still under study, 
and there is no universal model with a theoretical support to investigate SASI.
However, it is reasonable to assume that the frequency of SASI is almost constant in SFHx, as reported by Kuroda \textit{et al.}~\cite{kuroda2016sasi}.

Figure~\ref{fig:testhht} shows the results of HHT of the test signal.
In this analysis, a low trial number, namely,
$N_{\mathrm{eemd}} = 10^3$ is applied since the results do not change
even with higher values of $N_{\mathrm{eemd}}$ for a simple signal like Eq.\eqref{eq:testsignal}.
Figure \ref{fig:testimf} shows that sine-Gaussian wave was compared to the IMF5,
and the fact that the amplitude of the IMF5 is smaller than the sine-Gaussian wave in Fig.~\ref{fig:testhsa} indicates 
that mode splitting has occurred.
Probably the IMF1-3, 6 are mostly noise components.
The components of the sine-Gaussian wave split in the IMF4.
From Fig.~\ref{fig:testhsa}, 
the IF does not have a physical meaning 
owing to the influence of noise and calculation error 
where the amplitude is small, 
and the frequency vibrates around 130 Hz 
owing to the influence of the white Gaussian noise, 
even where the amplitude is large.
It is assumed that there are many cases 
where the mode splitting cannot be completely avoided in the actual analysis. 
In such cases, the analysis is advanced using these results.

\subsubsection*{\textbf{1. Estimating the starting point of the target mode}}
We estimated the appearance time $t_{\rm start}$ and disappearance time $t_{\rm end}$ of the signal on the IMF.
For this, we referred to~\cite{sakai2017estimation}, which used the root mean squared error (RMSE) and assumed a constant IF.

The appearance and disappearance times were defined using discrete time series signals as:
	\begin{align}
		t_{\rm start}(n_0)    &= n_0 \Delta t,\\
 		t_{\rm end}(n_0, N_T) &= t_{\rm start} + T \nonumber\\
		&= (n_0 + N_T)\Delta t,
	\end{align}
where, $n_0$ is a segment of $t_{\rm start}$,
$\Delta t$ is the sampling time interval, that is, inverse of the sampling frequency $(\Delta t=1/f_{\rm sampling})$,
$T$ is time of appearance of the signal, and
$N_T$ is the number of segments of $T$.
The segment [$n_0,n_0+N_T$] is estimated using $n_0$ and $N_T$ as variables.
		
\begin{figure}[ht]
        \centering
            \includegraphics[width=0.9\linewidth]{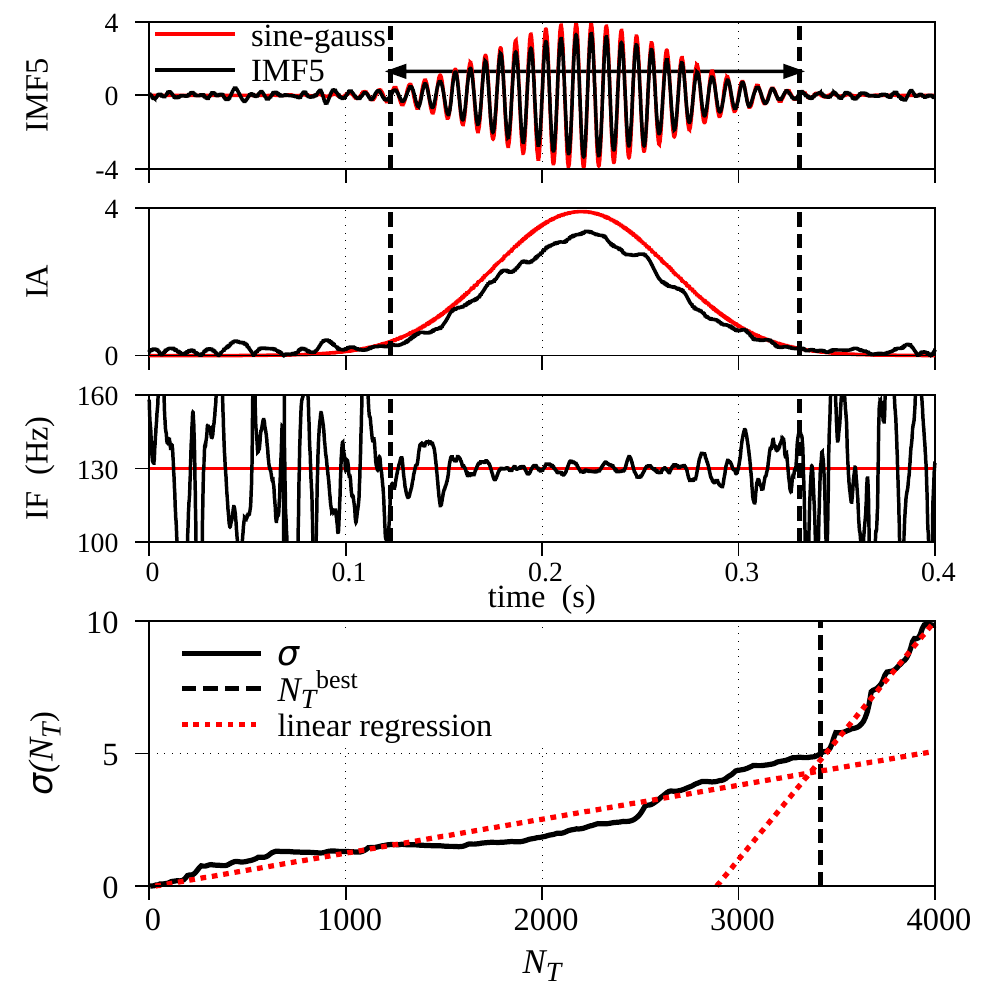}
            \caption{$N-\sigma$ map of the test signal}
             \label{fig:map-test}
    \end{figure}

The first step is to estimate the optimal $t_\mathrm{start}$ for each $N_T$. We calculate the IF variances $\hat{\sigma}(n_0, N_T)$ of the segment [$n_0, n_0 + N_T$] in all possible $n_0$ for a fixed $N_T$.
		\begin{align}
		  	\langle f \rangle (n_0, N_T) &= 
			\sum_{i=n_0}^{n_0+N_T-1} f_i w_i,
		\label{eq:meanf}
		\end{align}
		\begin{align}
			\mathrm{RMSE:}\ \ &\hat{\sigma}(n_0, N_T) \notag \\
			& = \sqrt{
			 	\frac{1}{N_T}
				\sum_{i=n_0}^{n_0+N_T-1}
				\big[f_i - \langle f \rangle(n_0, N_T)\big]^2
			},
		\end{align}
		where $f_i$ is IF, $a_i$ is IA, and
		$w_i$ is the weight function
                defined by
                \begin{equation}
                  w_i = \frac{a_i^2}{\displaystyle \sum_{i=n_0}^{n_0+N_T-1} a_i^2}.
                \end{equation}
                The optimal $n_0$ is represent by $n^\mathrm{best}_0$, with the smallest $\hat{\sigma}$ in the $N_T$.
		\begin{align}
			n^{\mathrm{best}}_0(N_T) &= 
				\mathop{\rm argmin}_{n_0}\ \hat{\sigma}(n_0,N_T),\\
			\sigma(N_T) &= 
				\hat{\sigma}(n^{\mathrm{best}}_0(N_T), N_T).
		\end{align}
		The lower part of Fig.~\ref{fig:map-test} shows the $N_T-\sigma$ map, which is obtained by performing the above calculations for all $N_T$.
		
The second step is to determine the optical $N_T$. We assume the following, 
			\begin{itemize}
				\item IF of the SASI is almost constant with time.
				\item The SASI mode appear in parts of the IMF.
			\end{itemize}
		On the $N_T-\sigma$ map, 
		$\sigma$ gradually increases $N_T$ 
		in the time segment where SASI dominates, 
		but it increases rapidly otherwise.
		Based on the above assumptions, 
		we divide the $N_T-\sigma$ domain into two regions: above and below N.
		Each region is then calculated by linear regression to determine the optimal $N_T^\mathrm{best}$ where the sum of both error terms is minimum.

			\begin{align}
				N_T^\mathrm{best} 
				= \mathop{\rm argmin}_{N_T}
				\big[
					&{\rm Er}(N_{T, \rm min},\ N_T) \notag\\
					&+ {\rm Er}(N_T+1,\ N_{T, \rm max})
				\big],
			\end{align}
			
			\begin{align}
				&{\rm Er}(N_{T,1}, N_{T,2})\notag\\
				&= \mathop{\rm min}_{a,b}
				\sqrt{
					\frac{\displaystyle \sum^{ N_{T,2} }_{ N_T=N_{T,1} }
							\Big[ \sigma (N_T)-(aN_T+b)
							\Big]^2
						}
						{ N_{T,2} - N_{T,1} }
				},
			\end{align}
			where $N_{T, 1}, N_{T, 2}\ (N_{T, 1} < N_{T, 2})$ denote any $N_T$ on the $N_T-\sigma$ map and
			$a, b$ denote the fitting coefficients, and
			$N_{T, \mathrm{min}}$ and $N_{T, \mathrm{max}}$ denote the minimum and maximum of the x axis on the $N_T-\sigma$ map, respectively.
			The black dashed line on the lower panel of Fig.~\ref{fig:map-test} 
			shows $N_T^\mathrm{best}$ and
			the black dashed line on the top panel 
			shows the time segment where SASI dominates.
	We succeeded in estimating the time segment 
	where the amplitude is not zero, 
	and this method is expected to function properly 
	in the time segment where the SASI mode dominates
	as shown in Fig.~\ref{fig:map-test}.




    \begin{figure}[ht]
        \centering
            \includegraphics[width=\linewidth]{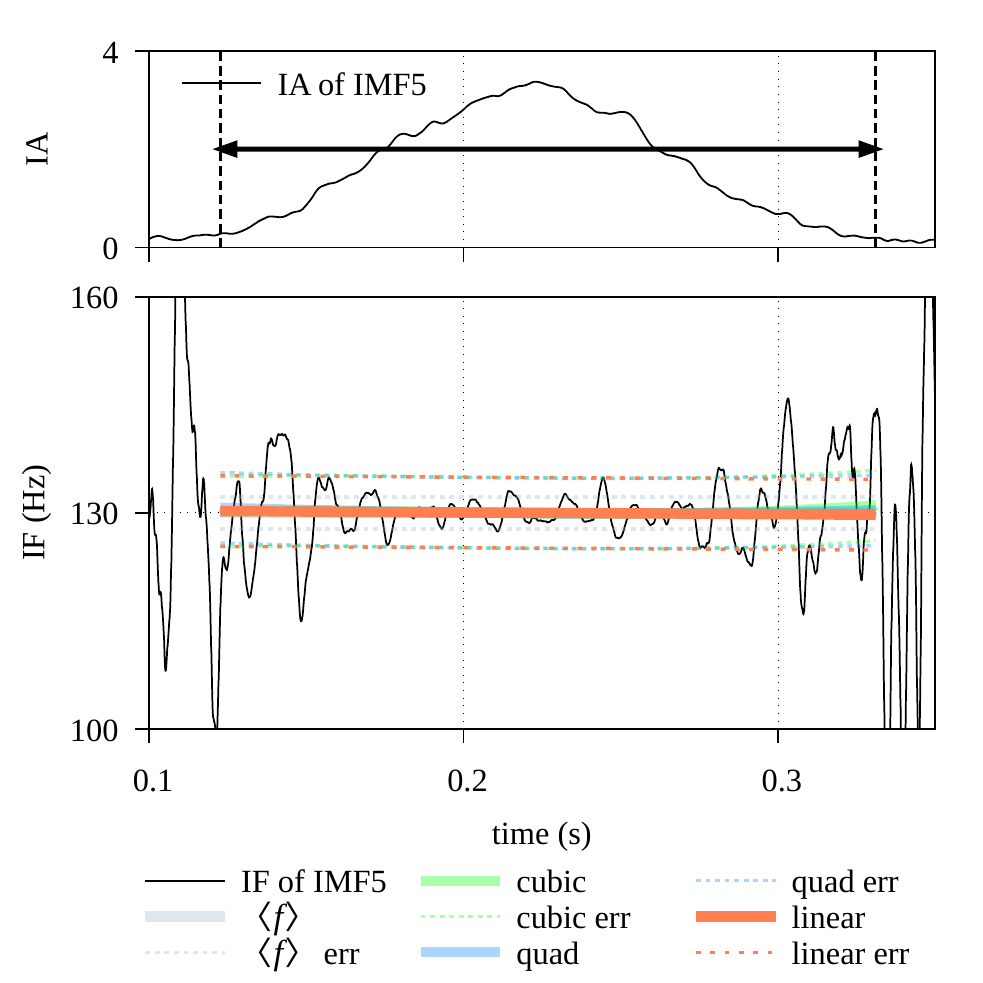}
            \caption{Fitting lines for the IF of the test signal}
            \label{fig:fit-test}
    \end{figure}

    \begin{table}
    \begin{ruledtabular}
    \caption{\label{tab:fit-test}Fitting coefficients and errors of the IF
            of the test signal. 
            The value in row of $\tau^I (I = 0, 1, 2, 3)$ are
            coefficients of $\tau^I$ and
            $f_\mathrm{err}^{\mathrm{fit}}$ is the error term of
            regression function. 
            The numbers after $\pm$ in the table 
            are the standard deviation of the coefficients, and $\chi^2$ is defined \eqref{eq:chi}.}
        \begin{tabular}{lrrr}
            &
            \multicolumn{1}{c}{\textrm{Linear}}&
            \multicolumn{1}{c}{\textrm{Quadratic}}&
            \multicolumn{1}{c}{\textrm{Cubic}} \\ 
            \hline
            $\tau^0$ &
            130.0 $\pm$ 1.0  &
            130.0 $\pm$ 1.2  &
            130.0 $\pm$\hspace{0.151cm} 1.3 \\
            $\tau^1$ &
            -0.3  $\pm$ 3.3  &
            -0.6  $\pm$ 3.5  &
            -0.4  $\pm$\hspace{0.151cm} 5.5 \\
            $\tau^2$ & &
            0.6   $\pm$ 8.0  &
            0.7   $\pm$\hspace{0.151cm} 8.3 \\
            $\tau^3$ & & &
            0.8   $\pm$ 16.8 \\      
            $f_{\rm err}^{\rm fit}$ &
             \multicolumn{1}{c}{4.9} &
             \multicolumn{1}{c}{4.9} &
             \multicolumn{1}{c}{6.5}  \\
            $\chi^2$  &
             $1.6 \times 10^{-4}$ &
             $4.4 \times 10^{-4}$ &
             $6.4 \times 10^{-4}$ \\
        \end{tabular} 
    \end{ruledtabular}
    \end{table}

	\begin{figure*}[pbth]
	    \centering
            \includegraphics[width=0.9\linewidth]{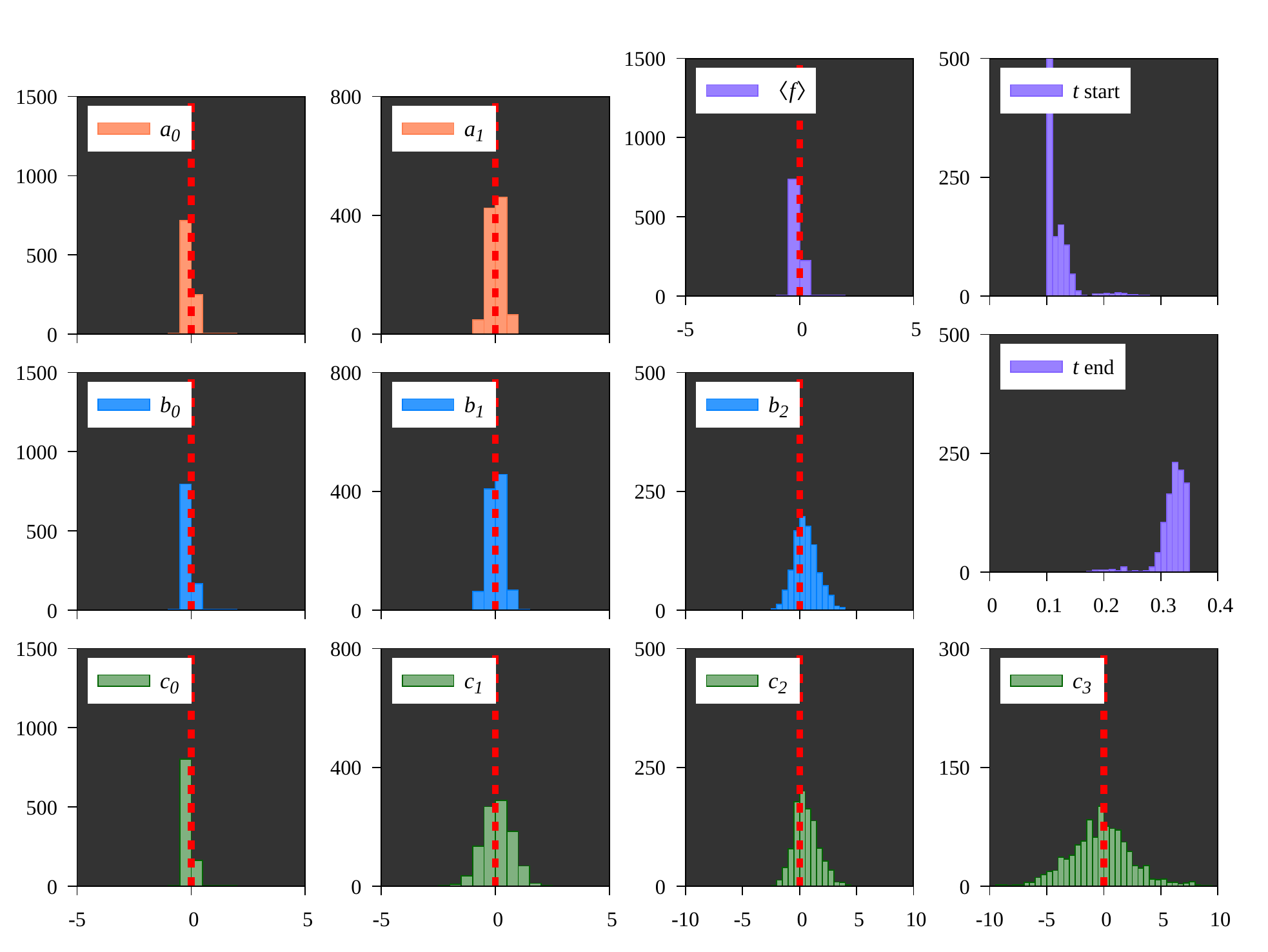}
            \caption{Histograms of the parameters estimated for the IF of the test signal. Here is a list of histograms of the fitting coefficients (Table \ref{tab:fit-test}), the average frequency, and the time interval.}
            \label{fig:hist-test}
    \end{figure*}
    
    

\subsubsection*{\textbf{2. Analysis of the frequency trend}}
As we mentioned in Sec.~\ref{sec:3DCCSNmodel}, there is something unclear about the SASI.
Therefore, it is necessary to obtain evidence from the observed data. Prior to that, we propose a method for analyzing the IF characteristics.
Here, the IF within the time segment, obtained by method (A), is analyzed.
We calculated the mean value of the IF in the trend using Eq.~\eqref{eq:meanf}, 
and used weighted least squares regression analysis for IF of the IMF to estimate the trend,
	\begin{align}
		f_{\rm lin}^{\rm fit}(t) &= 
			a_0 + a_1\tau 
			+ f_{\rm lin, err}^{\rm fit},\\
		f_{\rm quad}^{\rm fit}(t) &= 
			b_0 + b_1\tau + b_2\tau^2 
			+ f_{\rm quad, err}^{\rm fit},\\
		f_{\rm cubic}^{\rm fit}(t) &= 
			c_0 + c_1\tau + c_2\tau^2 + c_3\tau^3 
			+ f_{\rm cubic, err}^{\rm fit},
	\end{align}
where 
$f_{d}^{\rm fit}, d=\{ \mathrm{lin, quad, cubic} \}$ are dependent variables of the frequency trend, 
$\tau = (t-t_c)/(t_{\rm end} - t_{\rm start}),\ 
\ t_c = (t_{\rm start} + t_{\rm end} )/2$, 
$f_{d, \textrm{err}}^{\rm fit}, d=\{ \mathrm{lin, quad, cubic} \}$ are the error terms, and
$f_{d, \textrm{err}}^{\rm fit} = \sum w_i (f_i - f_{d}^{\rm fit})^2$. 
For regression analysis, we use the GNU Scientific Library~\footnote{\url{https://www.gnu.org/software/gsl/}} to calculate the best-fit coefficients and statistical errors.

In addition, the following values are introduced to investigate the accuracy of the analysis results,
	\begin{align}
		\label{eq:chi}
		\chi^2_d &= 
				\frac{1}{N_T}
				\sum_i \frac{ \big(f_{{\rm d},i}^{\rm fit} - f_{\mathrm{true},i} \big)^2 }{ f_{\mathrm{true},i} },\\
		d&=\{ \mathrm{lin, quad, cubic} \}, \notag
	\end{align}
where $ f_\mathrm{true} $ is the IF obtained by the HSA of the signal on which a white Gaussian noise is not superimposed and $ f_\mathrm{true} = 130 \unit{Hz}$.

The fitting result is shown in Fig.~\ref{fig:fit-test} and the coefficients are listed in Table~\ref{tab:fit-test}.
These results indicate 
that the test signal with a constant frequency at 130 Hz 
can be analyzed properly.
Additionally, 
when the weighting function is not used ($w_i = 1$), 
the average IF deviates from its original value;  
therefore, the weighting function is necessary to obtain an appropriate result.

\subsubsection*{\textbf{3. Error estimation}}
To estimate the error caused by the above method,
we prepared $N_\mathrm{test}=1000$ test signals 
and generated a histogram of the estimated values.

Figure~\ref{fig:hist-test} shows the histograms of
the fitting coefficients, the average frequency, and the time interval.
The top two histograms in the rightmost column are the results for the time interval. 
The top histogram in the the second column from the right is the result of the average frequency.
The rest are histograms of the fitting coefficients.
The widths of the histograms reflect variations in the results for $N_\mathrm{test}=1000$ test signals.
These are due to the fact that mode decomposition in EEMD could not be performed perfectly, and the noise behavior in which only the seed of Eq.~\eqref{eq:testsignal} in the second term appears different.
When applying this method to the GWs, the above should be considered a systematic error.

    \begin{figure}[pbht]
        \centering
            \includegraphics[width=0.9\linewidth]{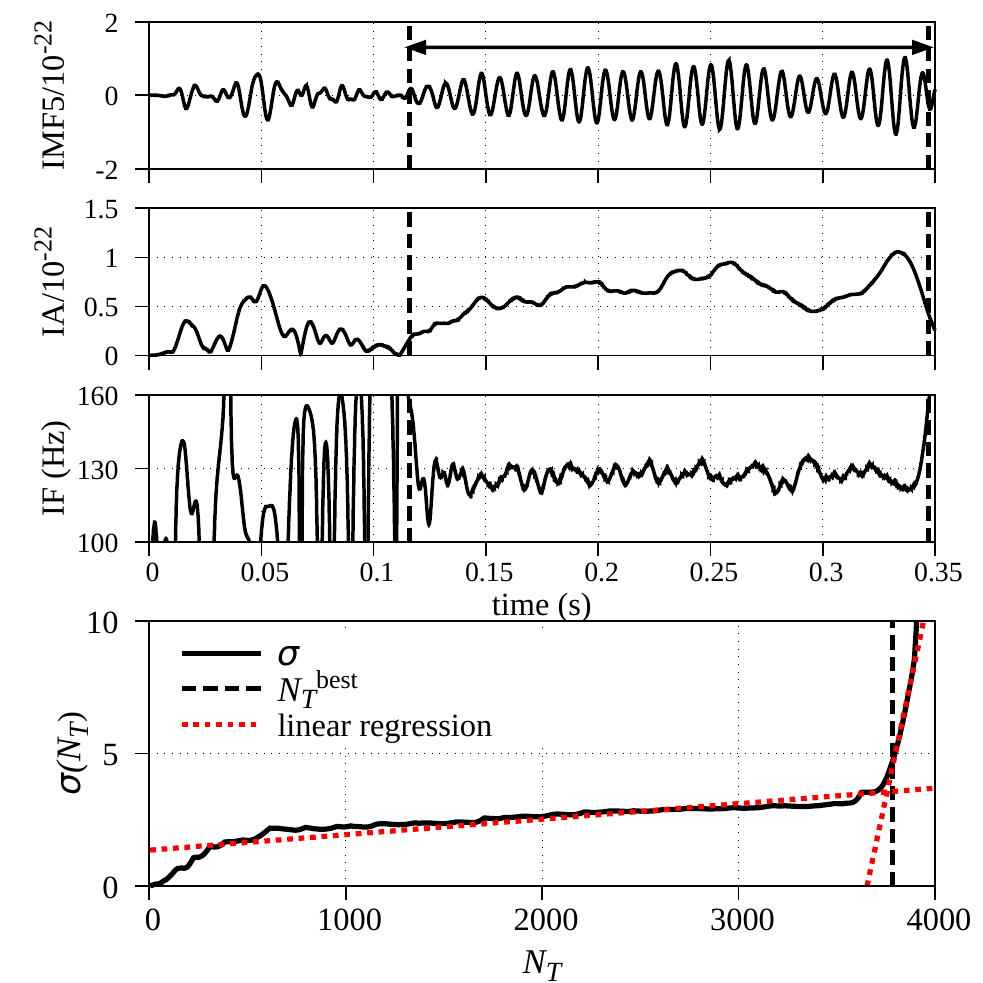}
            \caption{$N-\sigma$ map of the GW}
             \label{fig:map-gw}
    \end{figure}

    \begin{figure}[pbht]
        \centering
            \includegraphics[width=\linewidth]{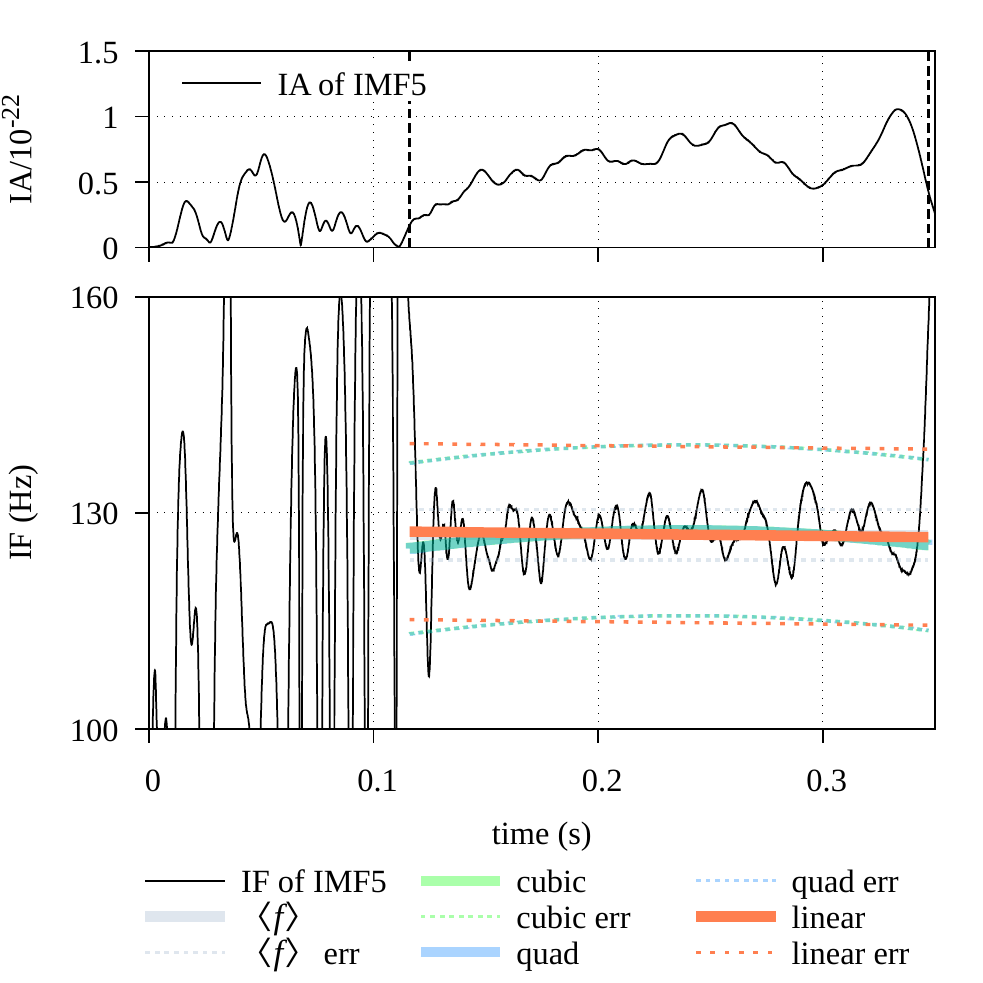}
            \caption{Fitting lines for the IF of the GW}
            \label{fig:fit-gw}
    \end{figure}

    \begin{table}[b]
    \caption{\label{tab:fit-gw}Same table as \ref{tab:fit-test} excluding
             $\chi^2$ for the simulated GW from the CCSN \cite{kuroda2016sasi}.}
    \begin{ruledtabular}
          \begin{tabular}{lrrr} 
                &
                \multicolumn{1}{c}{\textrm{Linear}}&
                \multicolumn{1}{c}{\textrm{Quadratic}}&
                \multicolumn{1}{c}{\textrm{Cubic}} \\ 
                \hline
                $\tau^0$ &
                127.0 $\pm$ 1.1 &
                127.6 $\pm$ 1.4  &
                127.6 $\pm$ 1.4 \\
                $\tau^1$ &
                -0.4  $\pm$ 2.0 &
                0.3   $\pm$ 2.2 &
                0.3   $\pm$ 4.8 \\
                $\tau^2$ & &
                -1.9  $\pm$ 3.9 &
                -2.3  $\pm$ 4.3 \\
                $\tau^3$ & & &
                -0.1  $\pm$ 8.4 \\      
                $f_{\rm err}^{\rm fit}$ &
                \multicolumn{1}{c}{12.2} &
                \multicolumn{1}{c}{11.9} &
                \multicolumn{1}{c}{11.9}  \\ 
            \end{tabular}
    \end{ruledtabular}
    \end{table}

	\begin{figure*}[pbth]
	    \centering
            \includegraphics[width=0.9\linewidth]{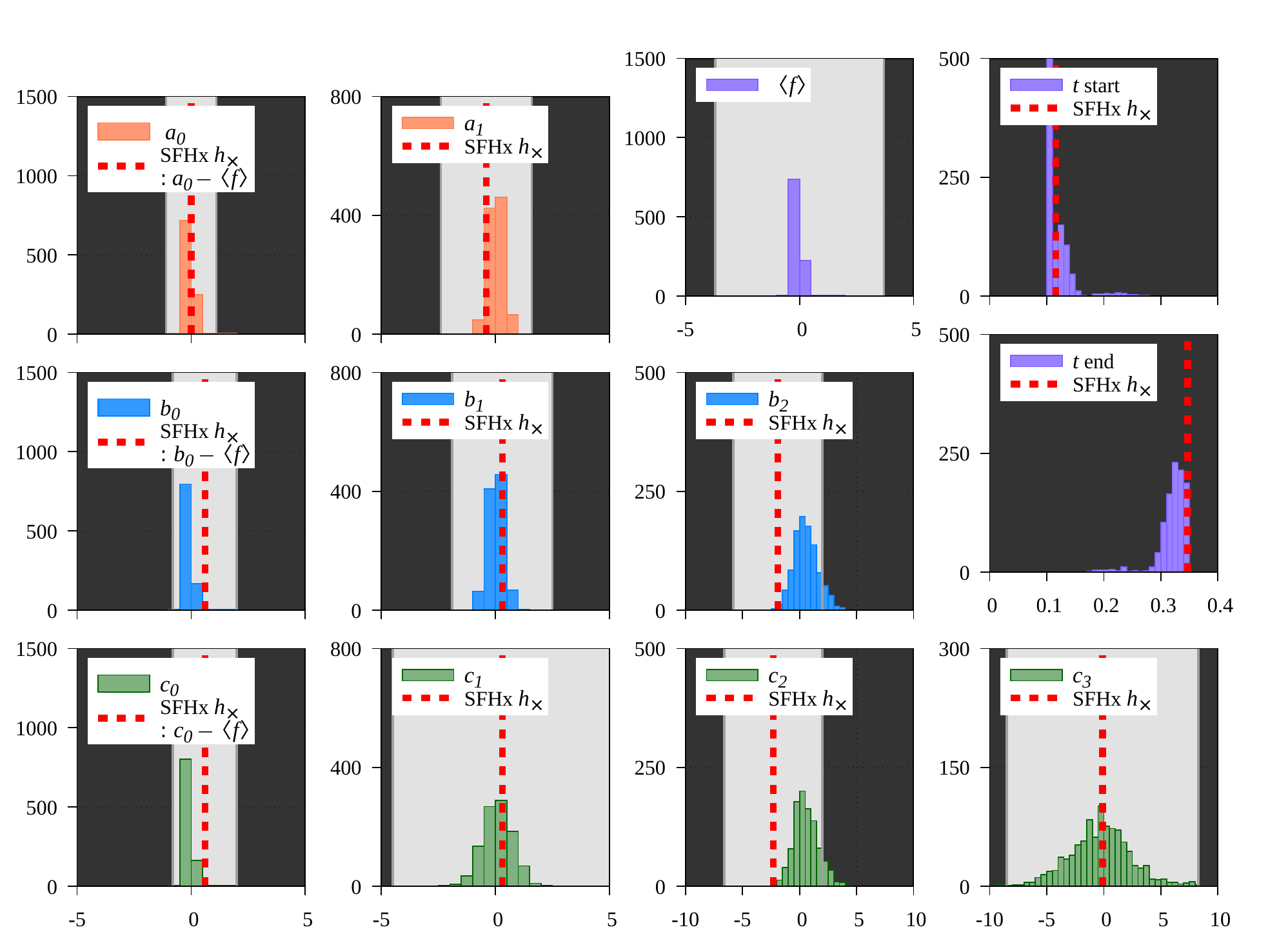}
            \caption{Estimated parameters of GW from the CCSN \cite{kuroda2016sasi}. The histograms of the test signal (Fig.~\ref{fig:hist-test}) and the parameters of the GW (Table~\ref{tab:fit-gw}) are shown together. The average frequency of the GW is shown in the error range around the 0 axis and is not shown a red dashed line. The time interval of the GW is shown as a red dashed line and is not shown in the error range.}
            \label{fig:hist-gw}
    \end{figure*}
    
    

\subsection{Analysis of the simulated GW from CCSNe}
We apply the above method to a simulated GW from a CCSN model of \cite{kuroda2016sasi} in Fig.~\ref{fig:iaifimf5}.
First, we estimated the time segment at which the SASI would appear.
Next, we searched for the optimal segment [$0.0 \unit{s},\ 0.35 \unit{s}$] to get $N_T^\mathrm{best}$. Using this, we further searched for [$0.0 \unit{s},\ N_T^\mathrm{best}$] and [$N_T^\mathrm{best},\ 0.35 \unit{s}$] to obtain the second and third $N_T^\mathrm{best}$, respectively.
We selected an $N_T^\mathrm{best}$ based on the smallest IF fitting error.
The results are shown in Fig.~\ref{fig:map-gw}.
We estimated that the SASI mode was dominant at $0.116 \unit{s} \lesssim t \lesssim 0.347 \unit{s}$ 
(t = time) and the internal environment that was needed to trigger SASI in the core was ready $\sim 0.1 \unit{s}$ after the core bounce ($t=0 \unit{s}$).
The end time ($\sim 0.35 \unit{s}$) is close to the time when the simulation ceased calculating, 
and it is speculated that the SASI may have temporarily weakened.
The fitting result is shown in Fig.~\ref{fig:fit-gw} and the coefficients are listed in Table~\ref{tab:fit-gw}.
Fig.~\ref{fig:hist-gw} shows the results of the GW analysis together with Fig.~\ref{fig:hist-test}, where the fitting coefficients are the red dashed lines and the error of the coefficients is the white range.
The average frequency of the GW is shown in the error range around the 0 axis and is not shown a red dashed line. The time interval of the GW is shown as a red dashed line and is not shown in the error range.
The values of $b_2$ and $c_2$ suggest that the frequency of SASI may fluctuate. However, considering the error generated in the test signal, this analysis method uses the GW obtained from the SASI mode in the segment. No time change was obtained from the IF.

\section{\label{sec:level5}Summary and discussion}
We analyzed a simulated GW signal from a core-collapse supernova (CCSN), which contained the strong SASI mode, using the HHT technique.
The frequency of the SASI has information on
the internal dynamics of the star before the explosion; 
therefore, we examined the behavior of the SASI frequency of the simulated signal.
The HHT method could isolate the IF of each physical mode that was generated in the signal source. 
Consequently, we endeavored to evaluate the feasibility of employing HHT to analyze GWs generated by the SASI mode.
Thus, we performed a statistical analysis because the IF was defined as a function of time, 
and found that the IF of the SASI mode extracted in the IMF5 was constant, namely $ \langle f \rangle = 127.0\pm 3.7 \unit{Hz}$, with time within $1\sigma$ in the time segment $0.116 \unit{s} \lesssim t \lesssim 0.347 \unit{s}$.


Our results were consistent with the previously reported research outcomes~\cite{kuroda2016sasi,kawahara2018linear} and established the effectiveness of the HHT technique.
Hence, we can expect the HHT to work well for analyzing the SASI mode from the observation signals, and it will be important to search for the signal that is superimposed on the simulated GW signals and GW detector noises.

Kuroda et al.~\cite{kuroda2016sasi} pointed out that the SASI mode can be captured by existing detectors (Advanced LIGO, Advanced Virgo, and KAGRA) if the event is in the Galaxy.
Therefore, there is a high possibility that SASI mode can be detected with Einstein Telescope~\cite{punturo2010einstein}, 
Cosmic Explorer~\cite{abbott2017exploring}, and thus, we plan to study the practical applications of HHT in the near future.

\begin{acknowledgments}
We thank K.~Hayama and T.~Takiwaki for their contribution in early phase of this work and for enlightening discussions.
We are also grateful to N.~Uchikata, R.~Nishi and K.~Watanabe for enlightening discussions. 
This work was supported in part by Japan Society for the Promotion of Science (JSPS) Grants-in-Aid for Scientific Research on Innovative Areas, Grant No. 24103005, No. JP17H01130,  No. JP17H06364, No. JP17H06358, and No. JP17H06361, by JSPS Core-to-Core Program A, Advanced Research Networks, and by the joint research program of the Institute for Cosmic Ray Research, University of Tokyo. This work was also supported in part by a Grant-in-Aid for JSPS Research Fellows [No. 20J00978 (S.~Tsuchida)], and by JSPS KAKENHI [Grant Nos. 17H06133 (N.~Kanda), No. 19K14717 and No. 21K13926 (K.~Sakai), and No. 19H0190 (H.~Takahashi)]. K.~Kotake was supported by Research Institute of
Stellar Explosive Phenomena at Fukuoka University and the associated project (No.207002).

\end{acknowledgments}



\bibliography{bib/CCSN,bib/KAGRA,bib/GW,bib/HHT}

\end{document}